\documentclass[twocolumn,showpacs,prl,superscriptaddress]{revtex4}

\usepackage{graphicx}
\usepackage{color}
\usepackage{tabularx}
\usepackage{epsfig}
\usepackage{amsmath}
\usepackage{amssymb}
\usepackage{graphicx}
\usepackage{dcolumn}
\usepackage{bm}
\usepackage{wasysym}
\usepackage{times}

\begin{document}

\title{Study of the de Almeida-Thouless line using
power-law diluted\\ one-dimensional Ising spin glasses}

\author{Helmut G.~Katzgraber}
\affiliation{Theoretische Physik, ETH Zurich, CH-8093 Zurich,
             Switzerland}
\affiliation{Department of Physics, Texas A\&M University, College Station,
             Texas 77843-4242, USA}

\author{Derek Larson}
\affiliation{Department of Physics, University of California, Santa Cruz,
	     California 95064, USA}

\author{A.~P.~Young}
\affiliation{Department of Physics, University of California, Santa Cruz, 
	     California 95064, USA}

\date{\today}

\begin{abstract}

We test for the existence of a spin-glass phase transition, the
de Almeida-Thouless line, in an externally-applied (random) magnetic
field by performing Monte Carlo simulations on a power-law diluted
one-dimensional Ising spin glass for very large system sizes.  We find
that an Almeida-Thouless line only occurs in the mean field regime,
which corresponds, for a short-range spin glass, to dimension $d$
larger than 6.

\end{abstract}

\maketitle

Perhaps the most surprising prediction of the mean-field theory of spin
glasses is that an Ising spin glass has a line of transitions in an
external magnetic field, the de Almeida-Thouless (AT)~\cite{almeida:78}
line. This instability line separates a high-temperature high-field
paramagnetic phase where relaxation times---possibly very large---stay
finite, from a low-temperature low-field phase where the energy
landscape has valleys separated by truly infinite barriers in the
thermodynamic limit. The AT line, an ergodic to non-ergodic transition
with no change in symmetry, is perhaps the most striking prediction
of the mean-field theory of spin glasses. Whether or not it occurs
in realistic systems is a major unsolved problem.

The existence or otherwise absence of an AT line in real (short-range)
spin glasses is also a key feature distinguishing the two most
popular scenarios for the nature of the spin-glass state below the
(zero-field) transition temperature: the replica-symmetry breaking
(RSB) picture of Parisi~\cite{parisi:80}, and the ``droplet picture''
of Fisher and Huse~\cite{fisher:86,fisher:88}. The RSB picture assumes
that the behavior of real spin glasses is very similar to that of the
mean-field solution~\cite{parisi:80} of the Sherrington-Kirkpatrick
infinite-range model.  Since the mean-field model has a stable
spin-glass state in a field and thus has an AT line, it is proposed
that this {\em also occurs for any short-range system} with a finite
temperature transition in zero field.  By contrast, the droplet
picture makes certain assumptions about the nature of the low-energy,
large-scale excitations (droplets) from which one finds {\em no AT
line in any dimension}.

Experimentally, it has been harder to determine if an AT line
occurs than to show that there is a transition in zero field. For
the latter case the divergence of the nonlinear susceptibility
provides a clear signature of the transition. Unfortunately, the
nonlinear susceptibility does not diverge in a field, i.e., along
the AT line.  However, as noted by two of us~\cite{young:04} there
is a closely-related static quantity which diverges on the AT line
and which can be measured in simulations, albeit not in experiments.
A finite-size scaling analysis of the two-point correlation length
indicated the absence of an AT line for three-dimensional (3D)
Ising spin glasses \cite{young:04,joerg:08a}.  Subsequently,
the same idea was applied to a one-dimensional (1D) model in
Ref.~\cite{katzgraber:05cKY} (referred to from now on as KY), in
which every spin interacts with every other spin in the system
with a strength which falls off with a power of the distance.
By varying the power, one can simulate the whole range of possible
behaviors~\cite{kotliar:83,fisher:88,katzgraber:05cKY}, from
infinite-range, through mean field, to non-mean field and finally to
the absence of a finite-temperature transition. This is analogous to
changing the space dimension $d$ of short-range finite-dimensional
models. KY found that an AT line does occur for parameter values
corresponding to the mean-field case (for short-range systems that
would be for $d \ge 6$), but not in the non-mean-field case ($d < 6$).
The possibility of a critical dimension above which the AT line
occurs had been considered before, see for example the discussion
in Ref.~\cite{newman:07}.

\paragraph*{Model and Observables.---}
\label{sec:model}

The model studied by KY is fully connected so the CPU time for
one Monte Carlo sweep (MCS) grows as ${\mathcal O}(L^2)$, where $L$
is the number of spins. This is inefficient for large $L$.  Recently,
this difficulty was removed in an elegant way in Ref.~\cite{leuzzi:08}
by diluting the interactions and fixing the connectivity $z$. We thus
study:
\begin{equation}
{\mathcal H} = -\sum_{i,j} \varepsilon_{ij} J_{ij} S_i S_j - \sum_i h_i S_i\; ,
\label{eq:hamiltonian}
\end{equation}
where $S_i = \pm 1$ are Ising spins evenly distributed on a ring of
length $L$ in order to ensure periodic boundary conditions. The sum is
over all spins on the chain and the couplings $J_{ij}$ are normally
distributed with zero mean and standard deviation unity (independent
of distance). 
The dilution matrix $\varepsilon_{ij}$ takes values $1$
or $0$, and a nonzero entry appears with probability $p_{ij}$, where
$p_{ij} \sim r_{ij}^{-2\sigma}$ with $r_{ij} = (L/\pi)\sin(\pi|i
- j|/L)$ representing the geometric distance between the spins.
The power $\sigma$ is a key parameter of the model. To avoid the
probability of placing a bond being larger than $1$, a short-distance
cutoff is applied and thus we take 
\begin{equation}
p_{ij} = 1 - \exp(-A/r_{ij}^{2\sigma}) \; ,
\;\;\;\;\;\;\;\;\;
z = \sum_{i = 1}^{L - 1} p_{iL} .
\label{eq:prob}
\end{equation}
The constant $A$ is determined numerically by fixing the average
coordination number $z$.  Note that this model has the same long range
interactions on average, $[J_{ij}^2]_{\rm av} \sim 1/r_{ij}^{2\sigma}$,
as in KY, but has only $Lz/2$ bonds rather than $L(L-1)/2$. Hence
the linear scaling of the CPU time for one MCS.

As in the fully-connected case \cite{katzgraber:05cKY}, by varying
$\sigma$ one can tune the model in Eq.~(\ref{eq:hamiltonian}) from the
infinite-range to the short-range universality class. For $0 < \sigma
\le 1/2$ the model is in the infinite-range universality class in
the sense that the parameter $A$ vanishes for $N \to \infty$, and for
$\sigma = 0$ it corresponds to the Viana-Bray model~\cite{viana:85}.
For $1/2 < \sigma \le 2/3$ the model describes a mean-field long-range
spin glass, corresponding---in the analogy with short-range
systems---to a short-range model in dimension above the upper
critical dimension $d \ge d_{\rm u} = 6$ \cite{katzgraber:08}. For
$2/3 < \sigma \le 1$ the model has non-mean-field critical behavior
with a finite transition temperature $T_c$. For $\sigma \ge 1$, the
transition temperature is zero.  We are interested in finite-range
models which have a non-zero $T_c$, i.e. $1/2 < \sigma \le 1$.

A rough correspondence between a value of $\sigma$ in the long-range
1D model and the value of a space dimension $d$ in a short-range
model can be obtained from
\begin{equation}
d = \frac{2 - \eta(d)}{2 \sigma - 1} \, 
\label{eq:d_sigma}
\end{equation}
where $\eta(d)$ is the critical exponent $\eta$ for the short-range
model, which is zero in the mean-field regime.  Equation
(\ref{eq:d_sigma}) has the following required properties (i) $d
\to\infty$ corresponds to $\sigma \to 1/2$, (ii) the upper critical
dimension $d_{\rm u} = 6$ corresponds to $\sigma_{\rm u} = 2/3$,
and (iii) the lower critical dimension, which is where $d_l
- 2 + \eta(d_l) = 0$, corresponds to $\sigma_l = 1$.
For example, in 3D, $\eta = 0.384(9)$ \cite{hasenbusch:08} and thus
the corresponding exponent is $\sigma \simeq 0.90$.

In this study we set the average coordination number to $z_{\rm av} =
6$ and use site-dependent random fields $h_i$ chosen from a Gaussian
distribution with zero mean $[h_i]_{\rm av} = 0$ and standard deviation
$[h_i^2]_{\rm av}^{1/2} = H_{\rm R}$. The latter has the advantage that
we can perform a detailed test for equilibration of the data when using
Gaussian-distributed interactions \cite{katzgraber:01,katzgraber:05cKY}
(see below).

To determine the existence of an AT line,
we compute the two-point finite-size correlation
length \cite{palassini:99b,ballesteros:00,young:04}. For this we start
by determining the wave-vector-dependent spin-glass susceptibility
given by
\begin{equation}
\chi_{\rm SG}(k) = \frac{1}{L} \sum_{i, j} \left[\Big(
\langle S_i S_j\rangle_T - \langle S_i \rangle_T \langle S_j\rangle_T
\Big)^2 \right]_{\rm av}\!\!\!\!\! e^{ik\, (i-j)} ,
\label{eq:chisg}
\end{equation}
where $\langle \cdots \rangle_T$ denotes a thermal average and
$[\cdots]_{\rm av}$ an average over the disorder. To avoid bias, each
thermal average is obtained from a separate copy of the spins, so we
simulate four copies at each temperature.
The correlation
length is given by \cite{katzgraber:05cKY}
\begin{equation}
\xi_L = \frac{1}{2 \sin (k_\mathrm{m}/2)}
\left[\frac{\chi_{\rm SG}(0)}{\chi_{\rm SG}(k_\mathrm{m})}
- 1\right]^{1/(2\sigma-1)} ,
\label{eq:xiL}
\end{equation}
where $k_\mathrm{m} = 2 \pi / L$ is the smallest non-zero wave-vector
compatible with the boundary conditions.  According to finite-size
scaling,
\begin{eqnarray}
&& {\xi_L}/{L} \sim {\mathcal X} [ L^{1/\nu} (T - T_c) ]  
\;,\;\;\;\;\;\;\; \sigma > 2/3
\nonumber \\
&& {\xi_L}/{L^{\nu/3}} \sim {\mathcal X} [ L^{1/3} (T - T_c) ] 
\;,\;\; 1/2 <\sigma \le 2/3,
\label{eq:xiscale}
\end{eqnarray}
with $\nu = 1/(2\sigma - 1)$ in the mean-field regime \cite{kotliar:83}.
Hence, if there is a transition at $T = T_c$,
data for $ {\xi_L}/{L}$ ($ {\xi_L}/{L^{\nu/3}}$ in the mean field region)
for different system sizes $L$ should cross at $T_c$.

We also present data for $\chi_{SG} \equiv \chi_{\rm SG}(0)$, which
has the finite-size scaling form
\begin{eqnarray}
&& \chi_{\rm SG} \sim L^{2 -\eta} {\mathcal C}[L^{1/\nu} (T - T_c)]
\;,\;\;\;\;\;\;\; \sigma > 2/3
\nonumber \\
&& \chi_{\rm SG} \sim L^{1/3} {\mathcal C}[L^{1/3} (T - T_c)]
\;,\;\; 1/2 <\sigma \le 2/3.
\label{eq:chisgscale}
\end{eqnarray}
Hence curves of $\chi_{\rm SG}/L^{2 - \eta}$ ($\chi_{\rm SG} /
L^{1/3}$ in the mean-field regime) should also intersect. This is
particularly useful for long-range models since $\eta$ is given by
the naive expression $2 - \eta = 2 \sigma - 1$ {\it exactly}.

As discussed in KY, for the simulations to be in equilibrium with
Gaussian fields and bonds, the following equality must hold:
\begin{equation}
U(\hat{q}_l, q) = - \frac{1}{T}\,\left[\frac{N_b}{L}\, (1 - \hat{q}_l)
		\right]_{\rm av} 
	        -\,  \frac{H^2_{\rm R}}{T}(1 - q) ,
\label{eq:energyqlh}
\end{equation}
where $q = L^{-1}\sum_i [\langle S_i \rangle^2_T]_{\rm av}$ is the
spin overlap, $\hat{q}_l =  N_b^{-1}  \sum_{i,j} \varepsilon_{ij}
\langle S_i S_j \rangle_T^2$ is the link overlap of a given sample,
and $N_b$ is the number of nonzero bonds of the sample.  To speed up
equilibration we use the parallel tempering (exchange) Monte Carlo
method \cite{geyer:91,hukushima:96}.  Simulations are performed at zero
field, as well as at $H_{\rm R} = 0.1$, a value considerably smaller
than $T_c(H_{\rm R} = 0)$ for the values of $\sigma$ studied. For
details see Table \ref{tab:simparams}.

\begin{figure*}[!tb]

\includegraphics[width=1\columnwidth]{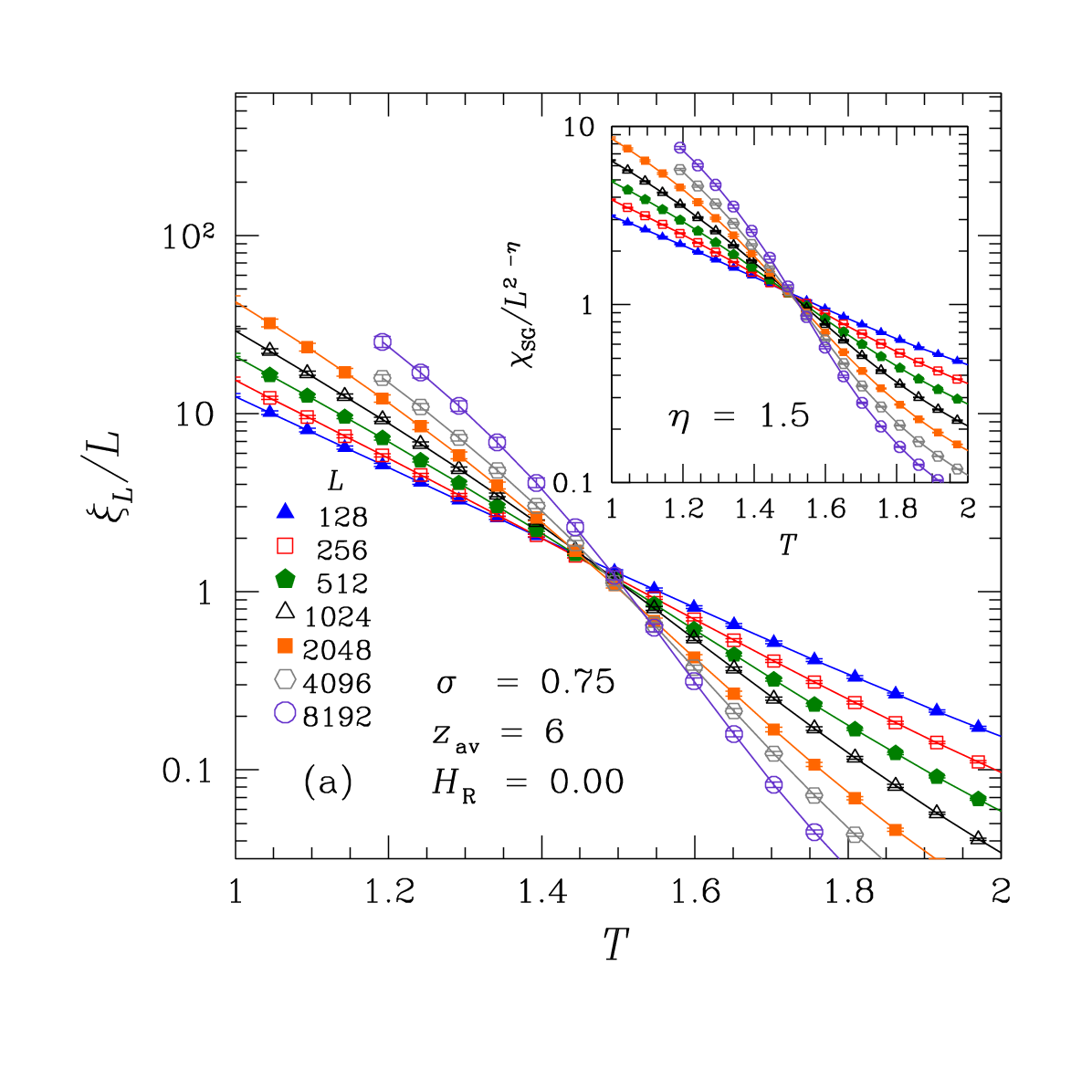}
\hspace*{-1.0cm}
\includegraphics[width=1\columnwidth]{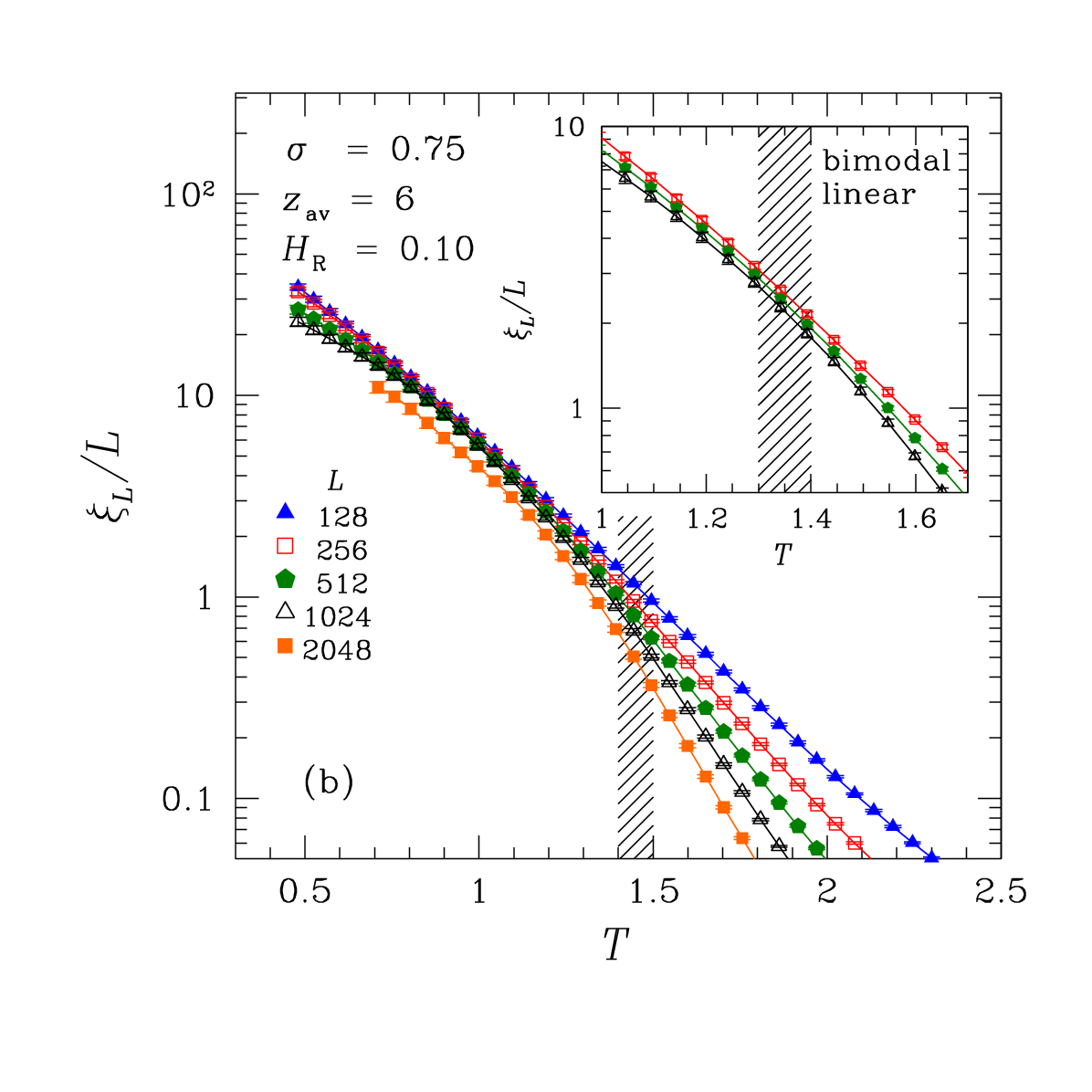}

\vspace*{-1.0cm}

\includegraphics[width=1\columnwidth]{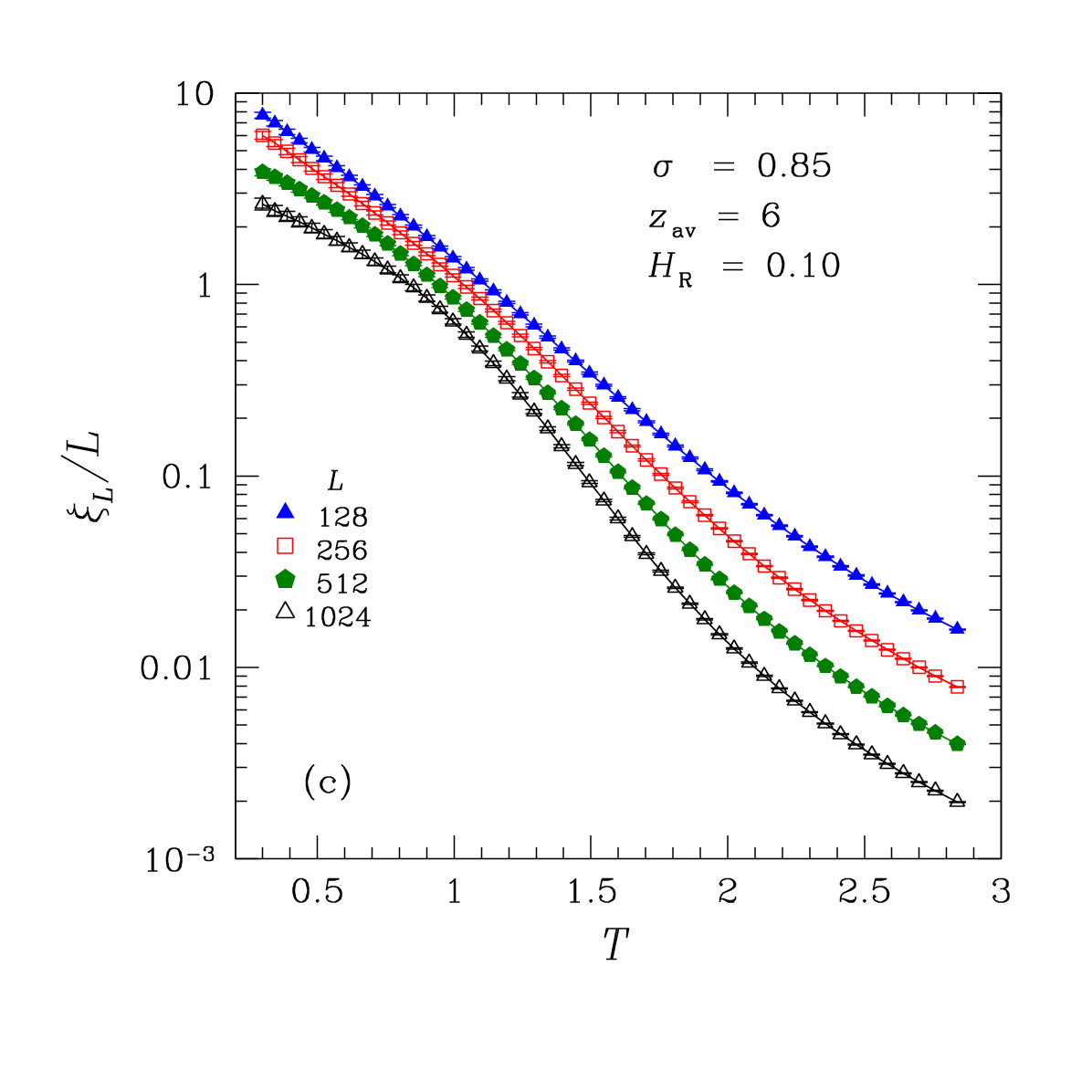}
\hspace*{-1.0cm}
\includegraphics[width=1\columnwidth]{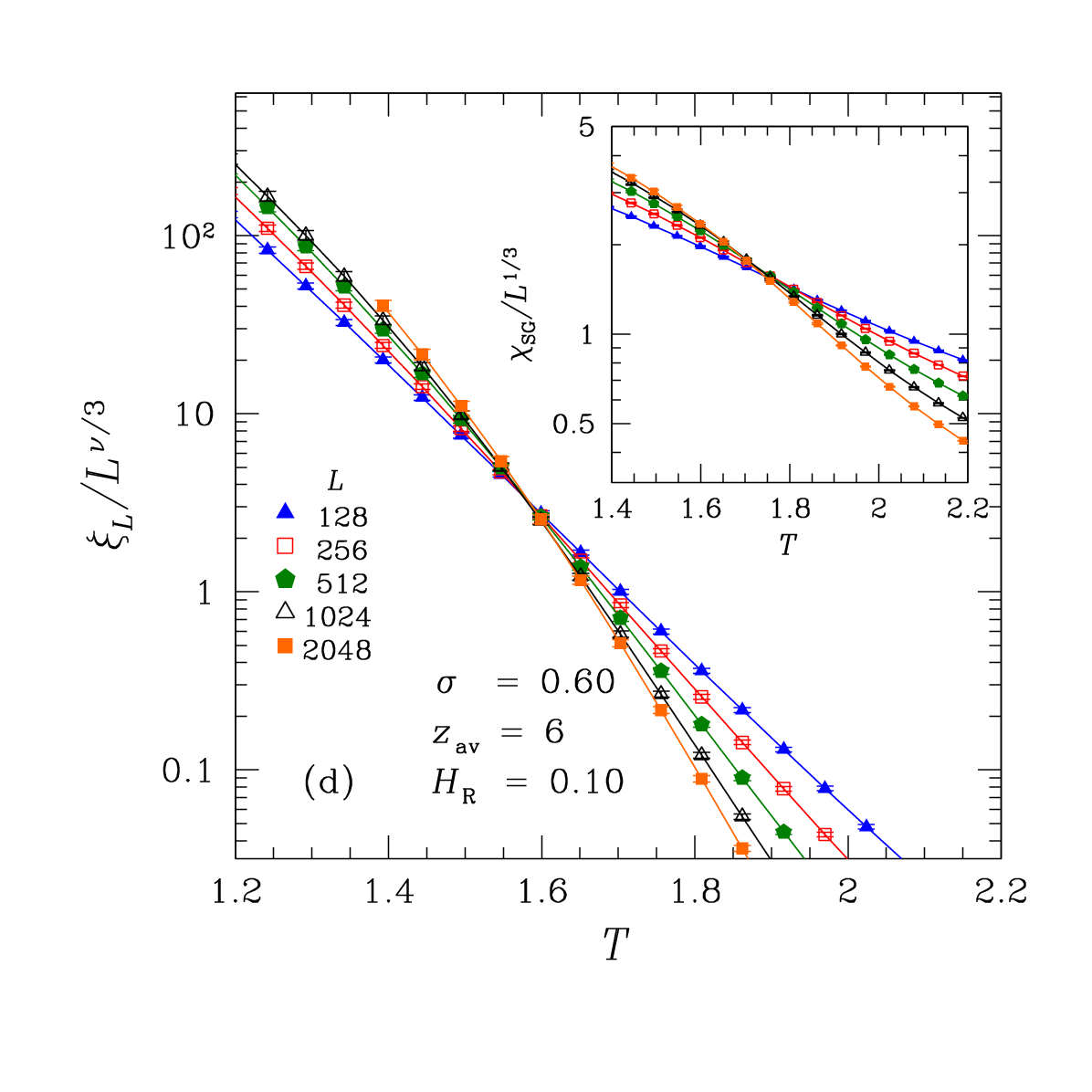}

\vspace*{-1.0cm}

\caption{(Color online)
Panel (a) Finite-size correlation length divided by $L$ as a function
of $T$ for different sizes for $H_{\rm R} = 0$ and $\sigma = 0.75$
(non-mean-field region).  The inset shows $\chi_{\rm SG}/L^{2-\eta}$
using the exact value $\eta = 3 - 2 \sigma = 1.5$. In both cases
the data cross indicating a phase transition at zero field. Panel
(b): Same as (a) but for $H_{\rm R} = 0.1$.  The absence of an
intersection down to low $T$ shows that there is no transition
in a field [the shaded area corresponds to $T_c(H_{\rm R} =
0)$]. The inset shows data for a bimodal ($\pm J$) distribution
of bonds, as used in Ref.~\cite{leuzzi:08b}, for sizes $L=256$
to $1024$ on a linear topology.  While Ref.~\cite{leuzzi:08b}
find a finite-temperature transition (shaded area in the inset)
we see no sign of it.  The absence of a transition is even more
clear in panel (c) where we show data as in (b) but for $\sigma =
0.85$, i.e., deeper into the non-mean-field regime. In panel (d)
we show data for the correlation length divided by $L^{\nu/3} \, (=
L^{5/3})$ as a function of $T$ for different sizes for $H_{\rm R}
= 0.1$ and $\sigma = 0.60$ (in the mean-field region).  The inset
shows $\chi_{\rm SG}/L^{1/3}$. The intersections show that there is
a transition in a field, i.e.,~an AT line for this value of $\sigma$.
}
\label{fig:results}
\end{figure*}

\begin{figure}[!tb]

\includegraphics[width=1\columnwidth]{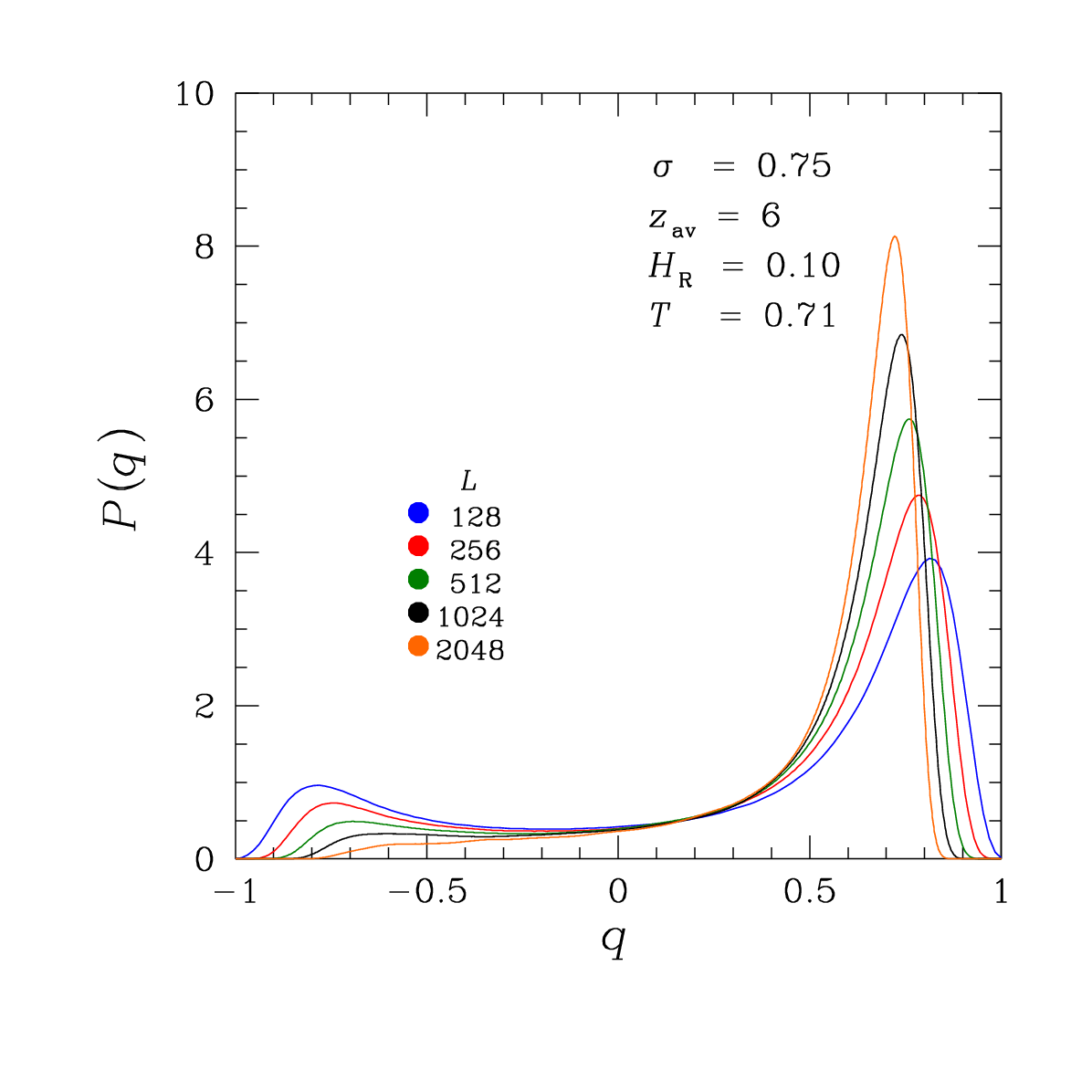}

\vspace*{-1.3cm}

\caption{(Color online)
Distribution of the spin overlap $q$ for $\sigma = 0.75$, $T = 0.71$
and $H_{\rm R} = 0.1$. Even for the largest $L$ studied there is a
tail which extends into the negative-$q$ region.
}
\label{fig:pq}
\end{figure}

\paragraph*{Results.---}
\label{sec:results}

We start by showing in Fig.~\ref{fig:results}(a) data for
$\xi_L/L$ against $T$ for $\sigma = 0.75$ in zero field, for several
system sizes. The data intersect cleanly at $T_c \simeq 1.50$
indicating a transition at that point, see Eq.~(\ref{eq:xiscale}).
The inset shows $\chi_{\rm SG} / L^{2-\eta}$ using the exact value
$\eta = 1.5$.

In contrast to Fig.~\ref{fig:results}(a), which shows the expected
zero-field transition for $\sigma = 0.75$, Fig.~\ref{fig:results}(b)
shows no intersections in a small field $H_{\rm R} = 0.1$
[approximately $0.067$ of the zero-field $T_c$ shown in
Fig.~\ref{fig:results}(a)].  Thus there is no AT line for $\sigma
= 0.75$, except possibly for even smaller values of the field.
Note that $\sigma = 0.75$ is in the non-mean-field regime ($2/3 <
\sigma < 1$). Whereas the data for $\sigma =0.75$ for small sizes
merge, and it is only for the {\em larger} sizes that the data do
not even meet, for $\sigma = 0.85$---deeper in the non-mean-field
regime---even the data for small sizes do not meet at any temperature
down to $T = 0.30$, see Fig.~\ref{fig:results}(c).

For comparison we also show data in the mean-field regime where an AT
line is expected to occur~\cite{katzgraber:05cKY}.
For $\sigma = 0.60$ and $H_{\rm R} = 0.1$
there is a clear intersection, see Fig.~\ref{fig:results}(d). The
temperature of the intersections is slightly different in the two
cases,  about 1.60 for $\xi_L/L^{5/3}$ and about 1.75 for $\chi_{\rm
SG}/L^{1/3}$,
suggesting finite-size effects, possibly due to long negative
tails in the spin overlap distribution; see Fig.~\ref{fig:pq}
and Ref.~\cite{leuzzi:08b}.

\begin{table}[!tb]
\caption{
Parameters of the simulations for different field strengths $H_{\rm R}$
and exponents $\sigma$. $N_{\rm sa}$ is the number of samples, $N_{\rm
sw}$ is the total number of Monte Carlo sweeps, $T_{\rm min}$ is the
lowest temperature simulated, and $N_T$ is the number of temperatures
used in the parallel tempering method for each system size $L$. The
last column shows the parameter $A$ [Eq.~(\ref{eq:prob})] fixing
$z_{\rm av} = 6$ neighbors.
\label{tab:simparams}}
{\footnotesize
\begin{tabular*}{\columnwidth}{@{\extracolsep{\fill}} c r r r r r r r}
\hline
\hline
$\sigma$ & $H_{\rm R}$ & $L$ & $N_{\rm sa}$ & $N_{\rm sw}$ & $T_{\rm min}$ &
$N_{T}$ & $A$
\\
\hline

$0.60$ & $0.10$ & $ 128$ & $8000$ &    $8192$ & $0.480$ & $46$ & $0.99458$\\
$0.60$ & $0.10$ & $ 256$ & $8000$ &   $32768$ & $0.480$ & $46$ & $0.90363$\\
$0.60$ & $0.10$ & $ 512$ & $5000$ &  $131072$ & $0.480$ & $46$ & $0.83827$\\
$0.60$ & $0.10$ & $1024$ & $5000$ &  $524288$ & $0.480$ & $46$ & $0.78926$\\
$0.60$ & $0.10$ & $2048$ & $4500$ &   $65536$ & $1.393$ & $26$ & $0.75140$
\\[2mm]

$0.75$ & $0.00$ & $ 128$ & $5000$ &   $32768$ & $0.300$ & $50$ & $1.71141$\\
$0.75$ & $0.00$ & $ 256$ & $5000$ &   $32768$ & $0.300$ & $50$ & $1.64289$\\
$0.75$ & $0.00$ & $ 512$ & $5000$ &  $524288$ & $0.300$ & $50$ & $1.59859$\\
$0.75$ & $0.00$ & $1024$ & $2900$ & $2097152$ & $0.300$ & $50$ & $1.56903$\\
$0.75$ & $0.00$ & $2048$ & $1000$ & $2097152$ & $0.480$ & $46$ & $1.54892$\\
$0.75$ & $0.00$ & $4096$ & $1000$ &   $65536$ & $1.192$ & $31$ & $1.53506$\\
$0.75$ & $0.00$ & $8192$ & $ 500$ &  $131072$ & $1.192$ & $31$ & $1.52544$ 
\\[2mm ]

$0.75$ & $0.10$ & $ 128$ & $5000$ &   $32768$ & $0.480$ & $46$ & $1.71141$\\
$0.75$ & $0.10$ & $ 256$ & $5000$ &  $131072$ & $0.480$ & $46$ & $1.64289$\\
$0.75$ & $0.10$ & $ 512$ & $5000$ &  $262144$ & $0.480$ & $46$ & $1.59859$\\
$0.75$ & $0.10$ & $1024$ & $5000$ &  $524288$ & $0.480$ & $46$ & $1.56903$\\
$0.75$ & $0.10$ & $2048$ & $2800$ &  $524288$ & $0.710$ & $39$ & $1.54892$\\
\\[2mm]

$0.85$ & $0.10$ & $ 128$ & $6000$ &   $16384$ & $0.300$ & $50$ & $2.39485$\\
$0.85$ & $0.10$ & $ 256$ & $6000$ &   $65536$ & $0.300$ & $50$ & $2.34867$\\
$0.85$ & $0.10$ & $ 512$ & $6800$ &  $524288$ & $0.300$ & $50$ & $2.32189$\\
$0.85$ & $0.10$ & $1024$ & $2500$ & $2097152$ & $0.300$ & $50$ & $2.30592$\\[2mm]

\hline
\hline
\end{tabular*}
}
\end{table}

We note that very recent work by Leuzzi {\em et al}.~\cite{leuzzi:08b}
comes to a different conclusion. Using Eq.~(\ref{eq:hamiltonian})
with bimodally-distributed disorder they find a transition in a
field in the non-mean-field regime, in particular for $\sigma =
0.75$ and $H_{\rm R} = 0.1$, where we do not find a transition,
see Fig.~\ref{fig:results}(b). We have no explanation for this
discrepancy. We have done several checks, including developing two
versions of the code independently and verifying that they give the
same results.  Furthermore, we have simulated the model with the
same bimodal disorder and geometry as used in Ref.~\cite{leuzzi:08b},
as well as the same field and $\sigma$ values, finding no signature
of a transition [see the inset to Fig.~\ref{fig:results}(b)].

\paragraph*{Summary.---}
\label{sec:summary}

Our conclusion, based on numerical results, is that there is an ``upper
critical dimension'' close to $6$ for the AT line. This agrees with
KY but disagrees with Ref.~\cite{leuzzi:08b}. This conclusion is
distinct from RSB theory \cite{parisi:80} which predicts {\em an AT
line in any space dimension with a zero-field transition}, and the
droplet picture \cite{fisher:86,fisher:88}, according to which there
is {\em no AT line in any finite dimension}.  Of course the numerical
data cannot rule out a transition at {\em extremely} small fields.

Note added in proof: We have recently heard (G. Parisi,
private communication) that there is an error in the analysis of
Ref.~\cite{leuzzi:08b}, and that their results for $\sigma = 0.75$
are now much more similar to ours.

\begin{acknowledgments}

We would like to thank T.~J\"org for discussions.
The authors acknowledge the Texas Advanced Computing Center (TACC)
at The University of Texas at Austin for providing HPC resources
(Ranger Sun Constellation Linux Cluster) and ETH Z\"urich for CPU
time on the Brutus cluster. They are also grateful to the Hierarchical
Systems Research Foundation for a generous allocation of computer time.
H.G.K.~acknowledges support from the Swiss National Science Foundation
under Grant No.~PP002-114713.

\end{acknowledgments}

\vspace*{-0.6cm}\bibliography{refs}

\end{document}